\documentclass[prd,twocolumn]{revtex4}
\usepackage{dcolumn}
\usepackage{multirow}
\usepackage{graphicx}
\usepackage{amssymb}
\usepackage{bm}
\usepackage{hyperref}%for .pdf
\usepackage{epstopdf}%for .pdf
\usepackage{color}
\usepackage{mathrsfs}
\usepackage{amsmath,amssymb,amsthm}
\usepackage{rotating}
\usepackage{sverb, longtable}
\usepackage{subfigure}

\usepackage[graphicx]{realboxes}
\usepackage{adjustbox}
\begin{document}

\title{Squeezing Cosmological Phase Transitions with International Pulsar Timing Array}
\author{Deng Wang}
\email{cstar@nao.cas.cn}
\affiliation{National Astronomical Observatories, Chinese Academy of Sciences, Beijing, 100012, China}
\begin{abstract}
A first-order MeV-scale cosmological phase transition (PT) can generate a peak in the power spectrum of stochastic gravitational wave background around nanohertz frequencies.
With the recent International Pulsar Timing Array data release two covering nanohertz frequencies, we search for such a phase transition signal. For the standard 4-parameter PT model, we obtain the PT temperature $T_\star\in$ [66 MeV, 30 GeV], which indicates that dark or QCD phase transitions occurring below 66 MeV have been ruled out at $2\,\sigma$ confidence level. This constraint is much tighter than $T_\star\sim$ [1 MeV, 100 GeV] from NANOGrav. 
We also give much tighter $2\,\sigma$ bounds on the PT duration $H_\star/\beta>0.1$, strength $\alpha_\star>0.39$ and friction $\eta<2.74$ than NANOGrav.
For the first time, we find a positive correlation between $\mathrm{log}_{10}T_\star$ and $\mathrm{log}_{10}H_\star/\beta$ implying that PT temperature increases with increasing bubble nucleation rate.
To avoid large theoretical uncertainties in calculating PT spectrum, we make bubble spectral shape parameters $a$, $b$, $c$ and four PT parameters free together, and confront this model with data. We find that pulsar timing is very sensitive to the parameter $a$, and give the first clear constraint $a=1.27_{-0.54}^{+0.71}$ at $1\,\sigma$ confidence level.

\end{abstract}
\maketitle

{\it Introduction.}---Since the discovery of gravitational waves (GW) by the LIGO interferometer \cite{LIGOScientific:2016aoc} at high frequencies, it is important and urgent to detect the stochastic gravitational wave background (SGWB) at low frequencies using pulsar timing arrays (PTA), which detect SGWB by the correlated deviations from time of arrivals (TOA) of radio pulses for a network of precisely timed millisecond pulsars within $\sim$ 1 kpc region from the earth. Currently, there are three independent PTA groups searching for SGWB using their long-term accumulated TOA data: (i) North American Nanohertz Observatory for Gravitational Waves (NANOGrav) \cite{Brazier:2019mmu}, Parkes Pulsar Timing Array (PPTA) \cite{Kerr:2020qdo} and European Pulsar
Timing Array (EPTA) \cite{Desvignes:2016yex}. The integration of these three groups forms the International Pulsar Timing Array (IPTA) \cite{Perera:2019sca}.       

Recently, it is exciting that the above four collaborations reported successively the strong evidence of a stochastic common spectrum process at low frequencies \cite{NANOGrav:2020bcs,Goncharov:2021oub,Chen:2021rqp,Antoniadis:2022pcn}, although PPTA group prefers discreetly identifying their result as an unknown systematic uncertainty \cite{Goncharov:2021oub}. However, there is no evidence found for a spatial correlation predicted by GW. In theory, GW sources of this possible SGWB detection can be interpreted as supermassive binary black holes (SMBBH) \cite{Vaskonen:2020lbd,DeLuca:2020agl, Kohri:2020qqd}, cosmological phase transitions \cite{Kosowsky:1992rz, Caprini:2010xv, Nakai:2020oit, Addazi:2020zcj, Ratzinger:2020koh, Li:2021qer}, cosmic strings \cite{Siemens:2006yp, Blanco-Pillado:2017rnf,Ellis:2020ena,Blasi:2020mfx}, domain wall decaying \cite{Hiramatsu:2013qaa}, large primordial curvature perturbations during inflation \cite{Kohri:2018awv}, and primordial magnetic field \cite{RoperPol:2022iel}. 

In this work, we explore the GW generated by the first-order cosmological phase transitions. As is well known, the electroweak PT temperature of the Standard Model in particle physics generally lies in the range $T_\star\lessapprox 100$ GeV. Nonetheless, phase transitions may take place at a relatively low temperature in the so-called hidden sectors \cite{Strassler:2006im, Chacko:2004ky, Schwaller:2015tja}. So far, there have been many cosmological, astrophysical and laboratorial probes to study the nature of hidden sectors \cite{Battaglieri:2017aum}. With the rapid development of observational techniques and gradual increasement of millisecond pulsars, PTA observations are demonstrated to have the ability to probe the dynamics of hidden sectors. Specifically, NANOGrav \cite{NANOGrav:2021flc} reported that the data can be explained with a strong first-order PT occurring below the electroweak scale and that a first-order PT is highly degenerated with SMBBH mergers as GW sources. PPTA \cite{Xue:2021gyq} found that pulsar timing is substantially sensitive to low-temperature PT lying in the range $T_\star\sim1-100$ MeV and can be used for constraining dark and QCD phase transitions. Other related studies that employ NANOGrav data to constrain phase transitions can be found in \cite{Ratzinger:2020koh, Moore:2021ibq}.

With the recent IPTA data release two (DR2) \cite{Chen:2021rqp}, which consists of 65 pulsars, we are dedicated to explore new properties of cosmological phase transitions. We find that IPTA DR2 can give much tighter constraints on PT parameters than NANOGrav and largely enhance our understanding of cosmological phase transitions. 

{\it Cosmological phase transitions.}---First order phase transitions take place via the locally tunneling of a field when there is a barrier between a true minimum and a false minimum of a potential. In the early universe, phase transitions are conducted by the nucleation of true vacuum bubbles, which expand over time in the background plasma. GW can be produced by collisions of a large number of bubbles and interacting bubble walls and background plasma.        

There are three main GW sources from phase transitions \cite{Weir:2017wfa, Breitbach:2018ddu} including bubble collisions \cite{Caprini:2007xq, Huber:2008hg}, collisions of sound wave originated from bubbles expansion \cite{Hindmarsh:2013xza,Hindmarsh:2016lnk}, and magnetohydrodynamics turbulence \cite{Caprini:2009yp} from bubbles expansion and sound wave collisions. As a consequence, total GW spectrum is expressed as $\Omega_{\mathrm{GW}}(f)=\Omega_{\mathrm{bub}}(f)+\Omega_{\mathrm{sw}}(f)+\Omega_{\mathrm{tur}}(f)$, where $f$ is the GW frequency. Taking reasonably parameterized models, the GW spectrum is written as \cite{Caprini:2009yp, Jinno:2016vai, Hindmarsh:2017gnf}    
\begin{equation}
h^2\Omega_{\mathrm{GW}}(f)=\mathcal{F}\Delta(v_w)S\left(\frac{f}{f^0_\star}\right)\left(\frac{\kappa\alpha_\star}{1+\alpha_\star}\right)^p\left(\frac{H_\star}{\beta}\right)^q,  \label{1}
\end{equation}
where $\mathcal{F}=7.69\times10^{-5}g_\star^{-\frac{1}{3}}$ explains the redshift of GW energy density, $g_\star$ is the number of relativistic degree of freedom when PT occurs, $\Delta(v_w)$ is a normalization factor depending on bubble wall velocity $v_w$, $\kappa$ is efficiency factor, $\alpha_\star$ denotes the PT strength that basically determines the amplitude of GW spectrum, $H_\star$ is the Hubble parameter at the PT temperature $T_\star$, $\beta$ is the inverse duration of PT, the function $S(f/f_\star^0)$ characterizes the spectral shape, where the present peak frequency $f_\star^0$ reads as 
\begin{equation}
f_\star^0\simeq 1.13\times10^{-10}\left(\frac{f_\star}{H_\star}\right)\left(\frac{T_\star}{\mathrm{MeV}}\right)\left(\frac{g_\star}{10}\right)^{\frac{1}{6}}\mathrm{Hz}, \label{2}
\end{equation} 
and $S(x)$ is described by three free parameters $a$, $b$, $c$ \cite{Jinno:2016vai}
\begin{equation}
S(x) = \frac{(a+b)^c}{(ax^{b/c}+bx^{-a/c})}.
\end{equation}

The values of normalization factor, efficient factor, peak frequency at emission $f_\star$, spectral shape, and two exponents $p$ and $q$ are shown in supplementary materials. $v_w$ and $\kappa$ are related to $\alpha_\star$ and the dimensionless fraction parameter $\eta$ \cite{Espinosa:2010hh}. 

To implement the constraints with IPTA DR2, we choose four models. The first is PT with four basic parameters $\{T_\star, \, \alpha_\star, \, H_\star/\beta, \, \eta\}$. Since the origin of SGWB may not just be cosmological PT, we take a GW source model which allows arbitrary overlapping contributions from PT and SMBBH. Furthermore, because PT GW spectrum is very sensitive to three bubble shape parameters, especially $a$, to avoid large theoretical uncertainty in analysis, we make $a$, $b$, $c$ and four PT parameters free in the third model. Consequently, we can explore more completely the parameter space of PT without the need of choosing specific values of $a$, $b$ and $c$ for envelope, semi-analytic or lattice simulation approaches..  
In order to study the integrated constraint on SGWB by combining CMB, BBN, and astrometry \cite{Smith:2006nka,Clarke:2020bil,Darling:2018hmc,Book:2010pf}, we also consider an uncorrelated common power-law (CPL) model, whose characteristic strain is $h_c(f)=A_{\mathrm{CPL}}(f/f_{\mathrm{yr}})^{(3-\gamma_{\mathrm{CPL}})/2}$ in the frequency range $f\in(f_l, f_h)$, where $A_{\mathrm{CPL}}$ and $\gamma_{\mathrm{CPL}}$ are amplitude and spectral slope. The integrated GW spectrum reads as $\tilde{\Omega}_{\mathrm{GW}}=\int_{f_l}^{f_h}(df/f)\Omega_{\mathrm{GW}}$.
We refer to the above four models as ``PTO'', ``PTBBH'', ``PTABC'' and ``CPL'', respectively.

\begin{figure}
	\centering
	\includegraphics[scale=0.45]{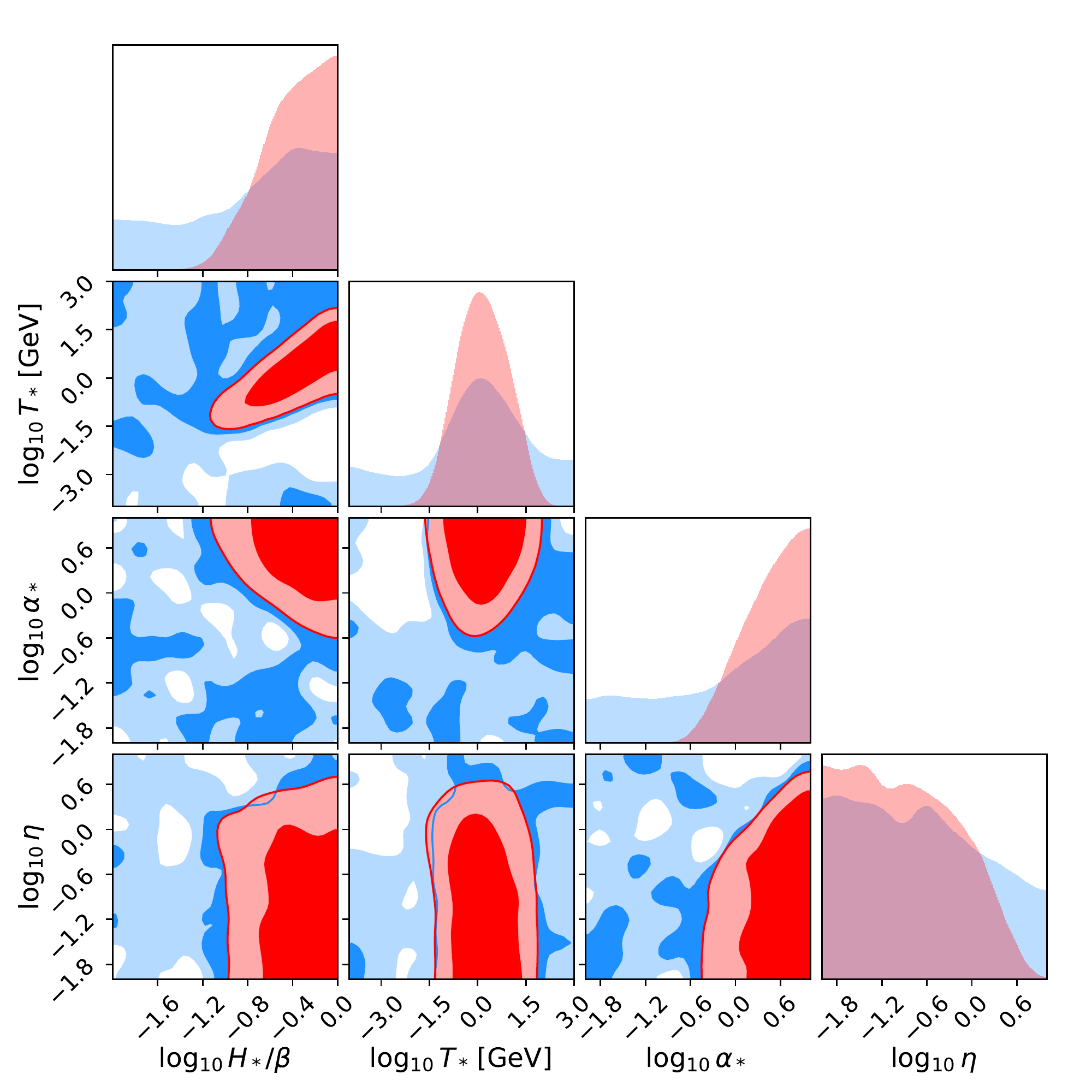}
	\caption{The marginalized posterior distributions of free parameters in PTO (red) and PTBBH (blue) models. We use the semi-analytic approach and choose the bubble shape parameters $a=1$, $b=2.61$ and $c=1.5$ for PTO and PTBBH models.}\label{f1}
\end{figure}

\begin{figure*}
	\centering
	\includegraphics[scale=0.55]{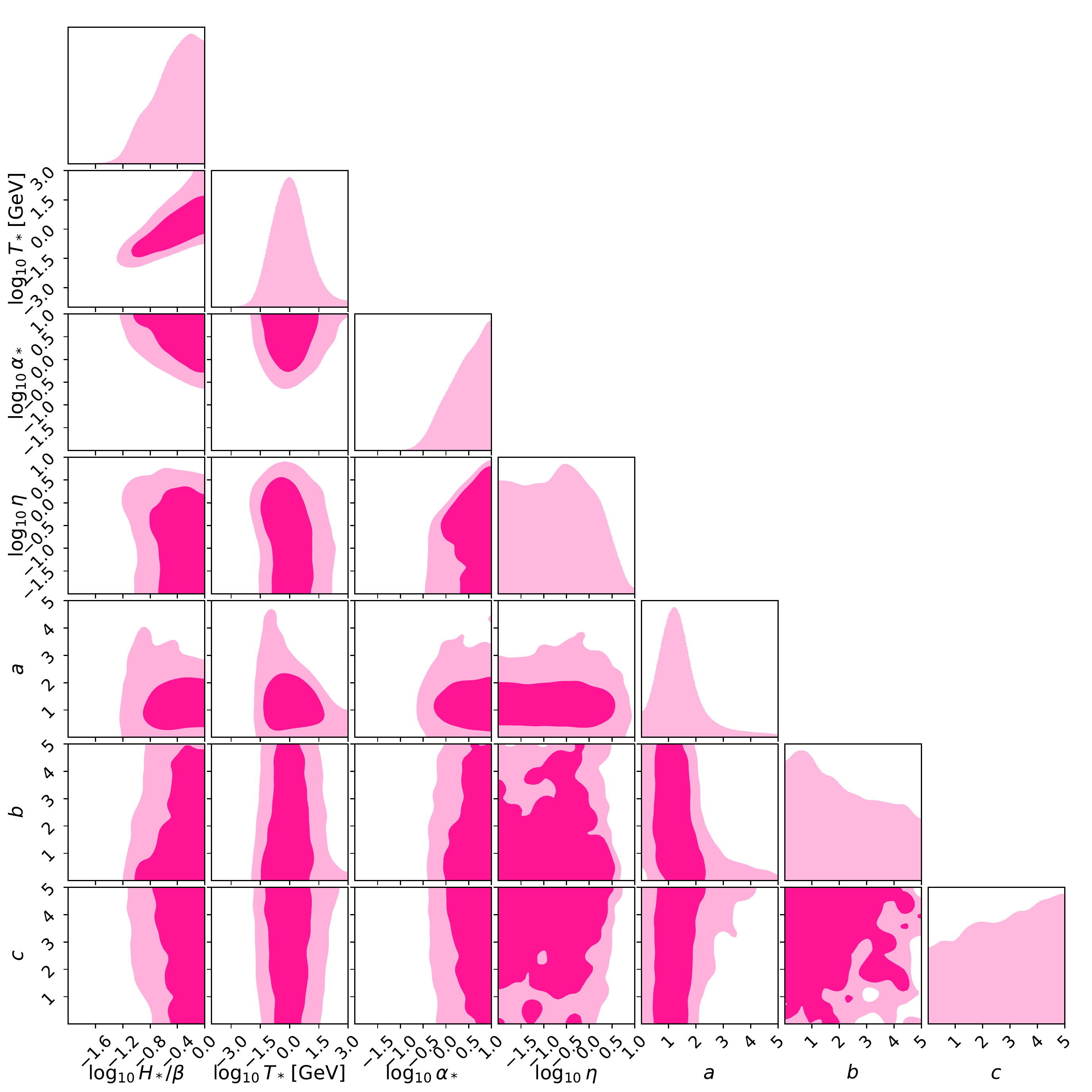}
	\caption{The marginalized posterior distributions of seven free parameters in the PTABC model. }\label{f2}
\end{figure*}

\begin{table*}
	\renewcommand\arraystretch{1.5}
	\caption{The $1\,\sigma$ (68\%) confidence ranges of free parameters and logarithmic Bayes factors for the PTO, PTBBH and PTABC models from IPTA DR2. To compute the Bayes factors, our reference model is CPL.}
	\setlength{\tabcolsep}{3mm}{
		{\begin{tabular}{@{}ccccccccc@{}} \toprule
				Parameters      &$\mathrm{log}_{10}H_\star/\beta$      &$\mathrm{log}_{10}T_\star$           & $\mathrm{log}_{10}\alpha_\star$                 &$\mathrm{log}_{10}\eta$           &$a$ &$b$ &$c$   & $\mathrm{ln}\,B_{ij}$                 \\ \colrule
				PTO       &$-0.38_{-0.35}^{+0.26}$  &  $0.12_{-0.79}^{+0.81}$    &$0.489_{-0.486}^{+0.362}$   &$-0.89_{-0.76}^{+0.81}$    &---    &---    &---     &0.09          \\  
				PTBBH           &$-0.71_{-0.85}^{+0.50}$  &$0.03_{-2.17}^{+1.49}$          &$-0.12_{-1.25}^{+0.79}$         &$-0.74_{-0.87}^{+1.04}$     &---       &---   &---          &0.99                        \\
				PTABC                  &$-0.42_{-0.39}^{+0.28}$       &$0.02_{-0.84}^{+0.85}$        &$0.500_{-0.503}^{+0.363}$   &$-0.70_{-0.89}^{+0.79}$  &$1.27_{-0.54}^{+0.71}$  &$2.01_{-1.42}^{+1.92}$         &$2.82_{-1.77}^{+1.52}$              &-0.72                        \\
				\botrule
			\end{tabular}
			\label{t1}}}
\end{table*}

\begin{figure}
	\centering
	\includegraphics[scale=0.6]{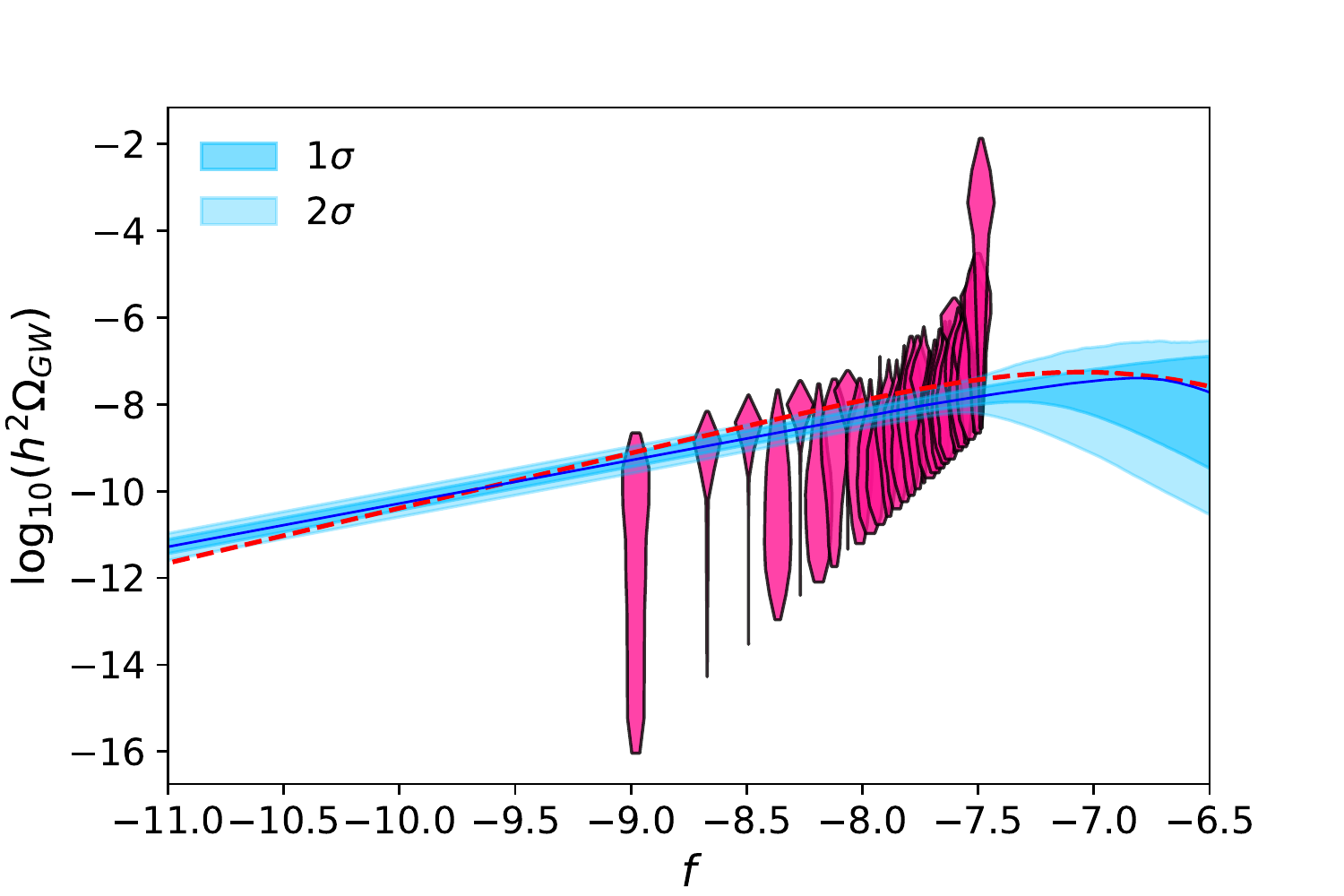}
	\caption{{\it The GW spectrum}. The pink violin plots are IPTA DR2 posterior distributions in 30 frequency bins. The solid blue and dashed red lines are median reconstructed spectrum from PTO and PTABC models, respectively. The shaded blue (light blue) band is $1\,\sigma$ ($2\,\sigma$) errors on PTO spectrum. }\label{f3}
\end{figure}

{\it Results.}---Using the IPTA DR2 posterior distributions on the delay spectrum, our marginalized constraints on PTO, PTBBH and PTABC models are presented in Figs.\ref{f1}-\ref{f2} and Tab.\ref{t1}. 
Comparing with NANOGrav's 12.5-yr analysis \cite{NANOGrav:2021flc}, to a large extent, we have compressed the PT parameter space. 
In Fig.\ref{f1}, for PTO, we obtain a tight constraint $\mathrm{log}_{10}T_\star=0.12_{-1.30}^{+1.34}$ at $2\,\sigma$ level, which indicates that the dark or QCD phase transitions occurring below 66 MeV are ruled out, and the permitted PT temperature range is [66 MeV, 30 GeV]. We find a positive correlation between $\mathrm{log}_{10}T_\star$ and $\mathrm{log}_{10}H_\star/\beta$ implying that PT temperature increases with increasing bubble nucleation rate. We obtain tighter lower bounds on PT duration $H_\star/\beta>0.1$ and strength $\alpha_\star>0.39$ and tighter upper bounds on the friction $\eta<2.74$ than NANOGrav's result \cite{NANOGrav:2021flc} at $2\,\sigma$ level. However, the $\alpha_\star$ and $\eta$ are still highly degenerate with left three parameters, respectively. Interestingly, for PTBBH, when considering simultaneously astrophysical foreground and cosmological background, the PT parameter space are substantially enlarged but the total tendency is consistent with the PTO case. The full 9-parameter contour for PTBBH are presented in supplementary material.    

For PTABC, we show the 7-dimensional parameter space in Fig.\ref{f2}. Very excitingly, IPTA data gives a strong constraint on bubble shape parameter $a=1.27_{-0.54}^{+0.71}$ at $1\,\sigma$ level. Due to the limited data quality, bubble shape parameters $a$, $b$ and $c$ are highly degenerated with four PT parameters. Similar to PTBBH, PT parameter space are also enlarged a little since the increasing number of free parameters. The corresponding constraining results are shown in Tab.\ref{t1}. In Fig.\ref{f3}, we find that the median reconstructed GW spectrum for PTABC lies outside the $2\,\sigma$ region of that for PTO around $10^{-8}$ and $10^{-11}$ Hz. This reveals that $a=1$, $b=2.61$ and $c=1.5$ is a slightly biased choice. We also observe that IPTA DR2 provides a very tight constraint on the evolution of GW abundance over time, and the extrapolated GW abundance $h^2\Omega_{\mathrm{GW}}=-11.28_{-0.16}^{+0.17}$ at $10^{-11}$ Hz, which is much more stringent than NANOGrav \cite{NANOGrav:2021flc}. To compare with NANOGrav better, we also show the reconstructed GW spectrum from first 5 frequencies in supplementary material.

In Fig.4, the confidence contour from IPTA DR2 \cite{Chen:2021rqp} is recovered and joint constraint on CPL parameter space are shown. We find that the inclusion of integrated bound $\tilde{\Omega}_{\mathrm{GW}}<10^{-6}$ helps rule out a large part of parameter space, i.e., many blue spectra permitted by IPTA, when considering high-frequency cutoffs $f_h=10^{-5}$ and $10^{-6}$ Hz. Three choices of low-frequency cutoffs $f_l=10^{-10}$, $10^{-12}$ and $10^{-10}$ Hz do not affect CPL parameter space from IPTA. 

In order to explore the observational viability of different GW source models, choosing CPL as the reference model, we compute the Bayesian evidence of each model, $\varepsilon_i$ and Bayes factor, $B_{ij}=\varepsilon_i/\varepsilon_j$, where $\varepsilon_j$ is the evidence of reference model. Following Ref.\cite{Trotta:2005ar}, we adopt a revised and conservative version of the so-called Jeffreys’ scale, i.e., $\mathrm{ln}\,B_{ij}=0-1$, $1-2.5$, $2.5-5$ and $>5$ indicate an {\it inconclusive}, {\it weak,} {\it moderate} and {\it strong} preference of the model $i$ relative to reference model $j$. For an experiment that leads to
$\mathrm{ln}\,B_{ij}<0$, it means the reference model is preferred by
data. Our results are shown in Tab.\ref{t1}. We find that there is no obvious preference between different models, except for PTBBH is approximately weakly preferred over CPL. Interestingly, PTBBH is weakly preferred over PTABC with an evidence of $\mathrm{ln}\,B_{ij}=1.71$.

\begin{figure}
	\centering
	\includegraphics[scale=0.6]{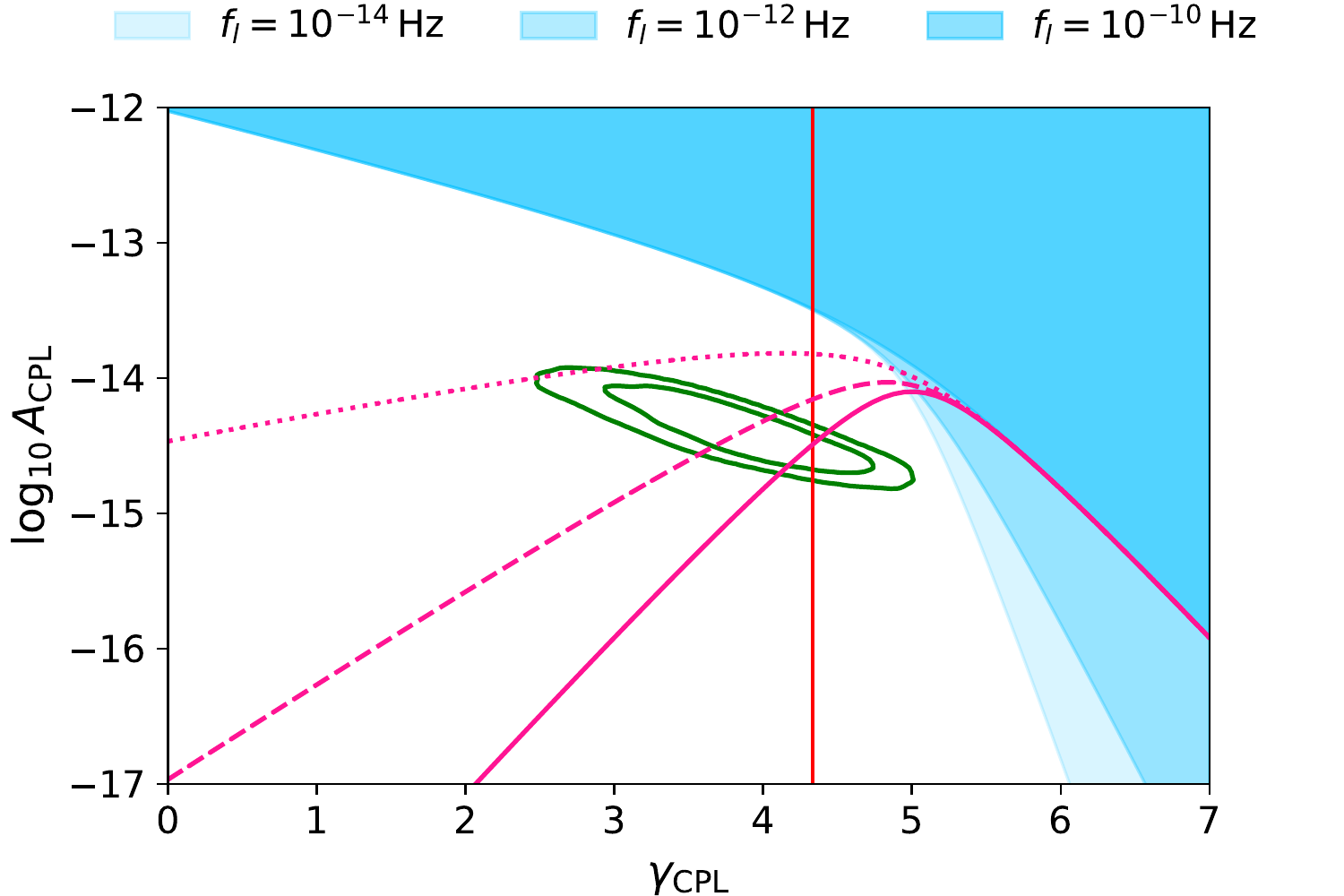}
	\caption{{\it The constraints on CPL GW spectrum.} The green contour is constraint from IPTA DR2 \cite{Chen:2021rqp}. The vertical line is theoretical prediction $\gamma_{\mathrm{CPL}}=13/3$ from SMBBH. For the joint constraint from CMB, BBN, astrometry and PTA, the shaded regions are excluded regions from low-frequency cutoffs, and the solid ($f_h=10^{-5}$ Hz), dashed ($f_h=10^{-6}$ Hz) and dotted ($f_h=10^{-7}$ Hz) lines are high-frequency cutoffs.}\label{f4}
\end{figure}

{\it Discussions and conclusions.}---With the recent IPTA DR2, we aim at searching for signals of cosmological phase transitions. Considering the simplest 4-parameter PT model, we obtain the permitted PT temperature $T_\star\in$ [66 MeV, 30 GeV], which indicates that the dark or QCD phase transitions occurring below 66 MeV have been ruled out by IPTA DR2 at $2\,\sigma$ confidence level. 
This constraint is much tighter than $T_\star\sim$ 1 MeV $-$ 100 GeV from NANOGrav \cite{NANOGrav:2021flc}. 
We also give tighter bounds on PT duration $H_\star/\beta>0.1$, strength $\alpha_\star>0.39$ and interaction between bubble wall and background plasma $\eta<2.74$ than NANOGrav \cite{NANOGrav:2021flc} at $2\,\sigma$ level.
These large improvements are originated from an efficient data combination of NANOGrav, EPTA and PPTA.  
For the first time, we observe a positive correlation between $\mathrm{log}_{10}T_\star$ and $\mathrm{log}_{10}H_\star/\beta$ implying that the higher PT temperature is, the higher the bubble nucleation rate is.
To avoid the large theoretical uncertainty in calculating PT spectrum, in the PTABC model, we make $a$, $b$, $c$ and four PT parameters free together, and for the first time, give a clear constraint on the bubble spectral shape parameter $a=1.27_{-0.54}^{+0.71}$ at $1\,\sigma$ confidence level. Considering cosmological phase transitions and SMBBH as GW sources simultaneously, we find the PT parameter space are substantially enlarged in the PTBBH model. Interestingly, different from PTBBH, 4-dimensional PT parameter space are just enlarged a little in PTABC. 
Since the median reconstructed GW spectrum for PTABC is inconsistent with   $2\,\sigma$ band from PTO reconstruction around $10^{-8}$ and $10^{-11}$ Hz, choosing $a=1$, $b=2.61$ and $c=1.5$ in the semi-analytic approach is obviously inappropriate. We argue that the best choice for a PTA experiment should be constraining directly PTABC and then reconstruct the corresponding PT spectrum, when searching for signals of cosmological phase transitions. 
IPTA DR2 provides a very tight constraint on the evolution of GW abundance over time, and the extrapolated GW abundance $h^2\Omega_{\mathrm{GW}}=-11.28_{-0.16}^{+0.17}$ at $10^{-11}$ Hz, which is much more stringent than NANOGrav. 
It is interesting that the inclusion of integrated bound $\tilde{\Omega}_{\mathrm{GW}}<10^{-6}$ helps rule out a large fraction of CPL parameter space, i.e., many blue spectra permitted by IPTA, when taking high-frequency cutoffs $f_h=10^{-5}$ and $10^{-6}$ Hz. In light of Bayes factor selection rule, we do not find any obvious preference between different models, except for PTBBH is approximately weakly preferred over CPL. Interestingly, PTBBH is weakly preferred over PTABC with an evidence of $\mathrm{ln}\,B_{ij}=1.71$. We believe that the next IPTA data release can significantly enhance our understanding of SGWB and cosmological phase transitions. 

{\it Acknowledgements.}---Deng Wang thanks Liang Gao, Jie Wang and Qi Guo for helpful discussions. This study is supported by National Nature Science Foundation of China under Grants No.11988101 and No.11851301. 
%Deng Wang acknowledges the usage of Numpy [], Pandas [], PTMCMC [] and Matplotlib [] during the preparation of this study. 

\end{document}